\documentclass[twocolumn,aps,floatfix,superscriptaddress,prl,showpacs]{revtex4}
\usepackage{amsmath, amsthm, amssymb, graphicx}
\input epsf.sty
\begin{document}

\def\g{\gamma}
\def\r{\rho}
\def\w{\omega}
\def\wo{\w_0}
\def\wp{\w_+}
\def\wm{\w_-}
\def\t{\tau}
\def\av#1{\langle#1\rangle}
\def\pf{P_{\rm F}}
\def\pr{P_{\rm R}}
\def\F#1{{\cal F}\left[#1\right]}

\title{Entropy production in nonequilibrium steady states:  A different approach and an exactly solvable canonical model}

\author{Daniel ben-Avraham}
\affiliation{Department of Physics, Clarkson University, Potsdam, NY 13699-5820, USA}
\affiliation{Department of Mathematics, Clarkson University, Potsdam, NY 13699-5815, USA}

\author{Sven Dorosz}
\affiliation{Theory of Soft Condensed Matter, Universit\'{e} du Luxembourg, Luxembourg, L-1511  Luxembourg}
%\affiliation{Department of Physics, Virginia Polytechnic Institute and State University, Blacksburg, Virginia 24061-0435, USA}

\author{Michel Pleimling}
\affiliation{Department of Physics, Virginia Tech, Blacksburg, Virginia 24061-0435, USA}

\date{\today}

\begin{abstract}
We discuss entropy production in nonequilibrium steady states by focusing on paths obtained by sampling at regular (small) intervals, instead of sampling on each change of the system's state.  This allows us to study directly entropy production in systems with microscopic  irreversibility,  for the first time.  
The two sampling methods are equivalent, otherwise, and  the fluctuation theorem holds also for the novel paths.
We focus on a fully irreversible three-state loop, as a canonical model of microscopic irreversibility, finding its entropy distribution,  rate of entropy production, and  large deviation function in  closed analytical form, and showing that the widely observed kink in the large deviation function arises solely from microscopic irreversibility. 

%We then analyze a simple system, consisting of a three-state loop, 
%and we obtain the entropy distribution, the rate of entropy production, and the large deviation function in a closed analytical form. This allows us to study the kink in the fluctuation function and other aspects of entropy production in great detail.  The same system is analyzed using the conventional path, and the results are contrasted with one another.
\end{abstract}
   
\maketitle

{\em Introduction.---} Entropy production is a hallmark of nonequilibrium steady states.  While entropy
production is a system-dependent quantity, there emerge some remarkable universal properties: for example, the probability distribution
of the total entropy production satisfies a detailed fluctuation theorem in large classes of systems (see, e.g., \cite{Eva93,Gal95,Kur98,Leb99,Sei05}), and a kink
appears in its large deviation function (and in that of related
currents) at zero entropy production \cite{Meh08,Vis06,Tur07,Cle07,Lac08,Mit10,Mal11}. Initially, this kink
has been attributed to specific properties of the systems under investigation, but a recent study indicates that it is a generic feature,
related to the detailed fluctuation theorem \cite{Dor11}.

Most published work on fluctuation theorems and the related large deviation functions deals with systems
that are {\em reversible} at the microscopic level: all transitions between states are bi-directional. However,  sheared granular systems
and chemical reactions where the products are cleared rapidly, are two of many important cases where microscopic reversibility is broken.  Few recent publications discuss 
fluctuation theorems for this type of systems.  Ohkubo has proposed a fluctuation theorem based on posterior probabilities \cite{Ohk09}, and
Chong {\it et al.} showed that an integral fluctuation theorem can be derived without  microscopic time reversibility~\cite{Cho10}.

Our aim in this  letter is two-fold. First, we propose the study of entropy production along
trajectories sampled at regular (small) intervals, instead of  the usual sampling on each change of the system's state. 
This novel sampling is equivalent to the traditional technique, in the limit of vanishingly small  intervals, and yields analogous results, including the  fluctuation theorems.  The advantage is that it enables direct analysis of systems with microscopic irreversibility, and  is more easily implemented in experiments and numerical studies. 
Second, we study the consequences of microscopic irreversibility by focusing on the smallest, canonical example: a fully irreversible three-state loop.  We thus find universal features of the entropy production and related quantities, and  demonstrate that the widely observed kink in the large deviation function at zero entropy is a certain
feature of irreversibility.

{\em Entropy production and two  kinds of sampling.---}
Consider a stochastic dynamical process in a system with a {\em discrete} set of states, $A, B, C,\dots$, and with transition rates $k(X,Y)$ (from state $X$ to $Y$).  We denote the steady-state probability of being in state $X$  by $\rho(X)$, and consider only systems with $\rho(X)>0$, for all states $X$. 

 {\em Event sampling}:  Imagine the system starting  from state $X_0$ (at the steady-state), and progressing through the sequence $X_1,X_2,\dots X_M$.  No other states occur between  $X_i$ and $X_{i+1}$.  The average time elapsed between two consecutive events is $\tau_i=1/k(X_i,X_{i+1})$.
This is the kind of trajectory, or path, employed in previous work on the subject
(see, e.g., \cite{Leb99,Lac08,Dor11,Bod10}).  

{\em Interval sampling}: We sample the system at $M$ regular intervals, $\tau, 2\tau,\dots, M\tau$, and record
its state  at each sampling, thus defining a trajectory $X_0,X_1,\dots,X_M$.   
The time gap between consecutive points on the trajectory is constant, $\tau_i=\tau$.  The system can be found in  the {\em same} state on  consecutive samplings, and it could also visit any number of states in between $X_i$
and $X_{i+1}$ \cite{BDP11}. One should note that this interval-sampling  is readily accessible in experiments, where one usually cannot record every transition between states,
as  would be needed for  event-sampling. 

The total entropy production, in the steady state, is given by \cite{Sei08}
\begin{equation}
\label{stot}
s_{tot}=\ln\frac{\rho(X_0)}{\rho(X_M)} + \ln\prod_i\frac{\w(X_{i-1},X_i)}{\w(X_i,X_{i-1})}\,,
\end{equation}
for either kind of trajectory.  For  interval sampling, $\w(X,Y)$ denotes the probability for finding the system in state $Y$, after time $\tau$, having started at state $X$ (at time zero).   For event sampling, $\w(X,Y)$ is replaced by $k(X,Y)$. 

If the sampling rate is large enough, $1/\tau\gg\max_{X,Y}k(X,Y)$, the most likely outcome for consecutive samplings is $X_i=X_{i+1}$,  and on the rare occasions that  $X_i\neq X_{i+1}$ no
other states are visited in between.  Repeated visits to the same state do not contribute to the entropy~(\ref{stot}), so    as
$\tau\to0$ interval sampling  becomes equivalent to event sampling.
Moreover, many of the properties found with the usual event sampling are reproduced by interval sampling, even for finite $\tau$.  For example, the detailed fluctuation theorem \cite{Leb99,Sei05},
$P(s_{tot})/P(-s_{tot})=\exp(-s_{tot})$, holds for both types of paths. 
A major advantage of interval sampling is that it lets us discuss situations of microscopic irreversibility: $X\to Y$, but $Y\not\to X$, and we focus on this idea.

{\em The 3-state loop.---}
For the sake of clarity, and for a chance at a full analytical solution, we wish to study the simplest nonequilibrium system
(with microscopic irreversibility).   A two-state system with non-trivial steady state (i.e., $\rho(A),\rho(B)>0$) is, per force,
an equilibrium system.  Thus, we are led to consider the 3-state system: $A\to B$, $B\to C$, $C\to A$, where we assume that all the rates are equal to $1$, thus defining our unit of time.  We later argue that despite its simplicity, this can be viewed as a canonical model for irreversibility.

Using the rate equations for the system, one finds,
\begin{equation}
\begin{split}
&\wo=\frac{1}{3}+\frac{2}{3}e^{-3\tau/2}\cos\left(\frac{\sqrt{3}}{2}\tau\right)\,,\\
&\wp=\frac{1}{3}+\frac{1}{3}e^{-3\tau/2}\left[-\cos\left(\frac{\sqrt{3}}{2}\tau\right)+\sqrt{3}\sin\left(\frac{\sqrt{3}}{2}\tau\right)\right]\,,\\
&\wm=\frac{1}{3}+\frac{1}{3}e^{-3\tau/2}\left[-\cos\left(\frac{\sqrt{3}}{2}\tau\right)-\sqrt{3}\sin\left(\frac{\sqrt{3}}{2}\tau\right)\right]\,,
\end{split}
\end{equation}
where $\wo\equiv\w(A,A)=\w(B,B)=\w(C,C)$ denotes the {\em neutral} transitions, $\wp\equiv\w(A,B)=\w(B,C)=\w(C,A)$
are the {\em forward} transitions, and $\wm\equiv\w(A,C)=\w(B,A)=\w(C,B)$
the {\em reverse} transitions.  Although these exact expressions can be employed in the subsequent calculations, we are interested in the limit $\tau\to0$, and in effect we use their lower-order expansions:
$\wo=1-\tau+\tau^2/2+\dots$, $\wp=\tau-\tau^2+\dots$, and $\wm=\tau^2/2-\t^3/2+\dots$.  We have verified carefully that
the final results are not affected.  Note that the ratio $\wm/\wp\approx\tau/2$, for the ``forbidden" reverse direction, vanishes as $\t\to0$.

{\em Probability distribution of entropy production.---}
Since $\rho(A)=\rho(B)=\rho(C)=1/3$, the first term on the rhs of (\ref{stot}) does not contribute to $s_{tot}$.  The remainder,
which we denote simply by $s$, is the entropy produced in the thermal bath coupled to our system.  Of the three types of terms that appear inside the product describing $s$, $\wo/\wo$, $\wp/\wm$, and $\wm/\wp$, only the last two contribute to $s$, in equal and opposite amounts.  Thus, $s$ assumes a 
discrete spectrum of values: $s_m=m\,ds$, with $ds = \ln(\wp/\wm)$ and $m=0,\pm1,\pm2,\dots,\pm M$,
%\begin{equation}
%s_m=m\,ds\,; \qquad ds = \ln(\wp/\wm)\,,\qquad m=0,\pm1,\pm2,\dots,\pm M\,,
%\end{equation}
where $m=N_+-N_-$ is the excess number of forward ($N_+$) over reverse  ($N_-$) transitions.  

\begin{figure}[h] 
%\centerline{\epsfxsize=4.00in\ \epsfbox{nu_fig.ps}}
\centerline{\epsfxsize=3.20in\ \epsfbox{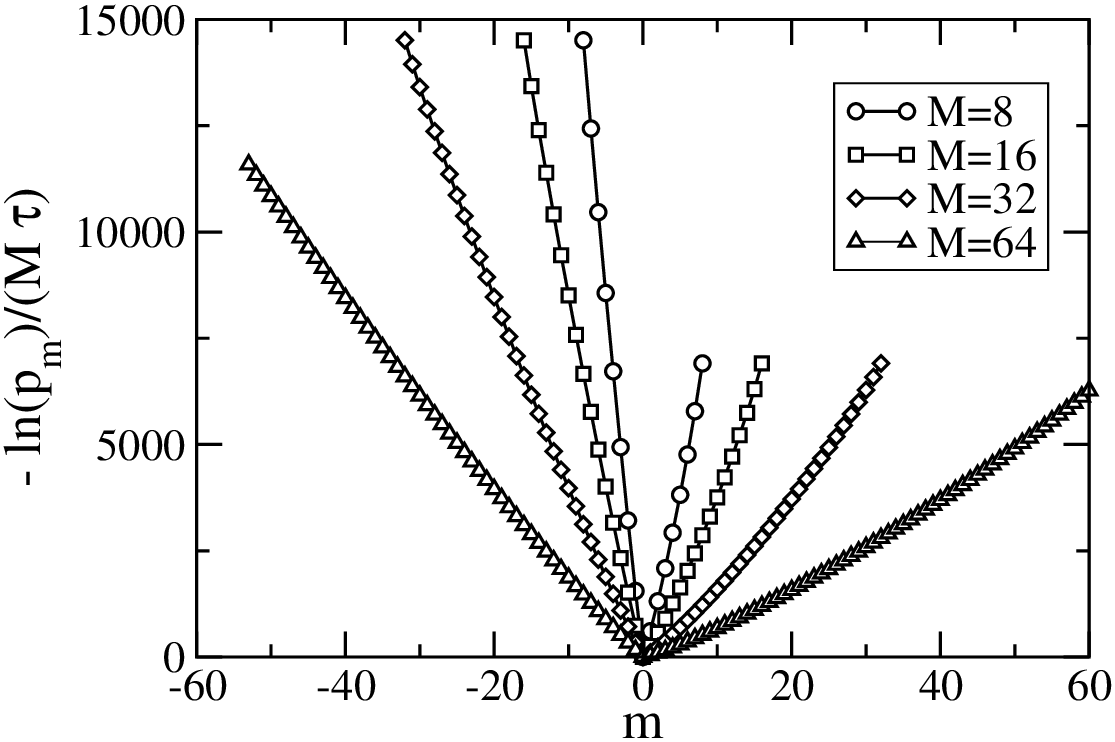}}
\caption{The probability $p_m$ as a function of $m$, the total number of forward and reverse transitions, for the 
3-state loop where paths are sampled over $M$ intervals of length $\tau = 0.001$.
In order to make the different curves obtained for different values of $M$ more easily distinguishable, we plot
$- \ln(p_m)/(M \tau)$ where $M\tau$ is the total length of the path. One should note the emergence of a kink for
increasing values of $M$.
}\label{fig1}
\end{figure}

The probability $p_m$ of
obtaining $s_m = m\,ds$, is the sum of the weights of all the trajectories consistent with that value.  The weight
of a trajectory with $N_+$-forward, $N_-$-reverse, and $N_0$-neutral transitions is
$\wo^{N_0}\wp^{N_+}\wm^{N_-}$.  All values of  $N_+,N_-,N_0$ must be counted, subject to the constraints $N_+-N_-=m$ and $N_++N_-+N_0=M$.
For any finite $M$, one can work out explicit (cumbersome) expressions.  Alternatively, the sums can be easily worked out numerically, see Fig.\ \ref{fig1}.

%%%%%%%%%%%%%%%%%%%%%%%%%%%%%%%%%%%%%%%%%%%FIG 1.%%%%%%%%%%%%%%%%%%%%%%%%%%%%%%%%%%%%%%%%%%%%%%%%%%%%%%
%%%%%%%%%%%%%%%%%%%%%%%%%%%%%%%%%%%%%%%%%%%FIG 1.%%%%%%%%%%%%%%%%%%%%%%%%%%%%%%%%%%%%%%%%%%%%%%%%%%%%%%

The problem can be approached more elegantly using the generating function $p(z)=\sum_m p_mz^m$.  In our case,
the generating function is clearly
\[
p(z)=\left(\wo+\wp z+\frac{\wm}{z}\right)^M\,,
\]
since its trinomial expansion yields all the possible combinations subject to the constraint $N_++N_-+N_0=M$, and the 
$z^m$-terms are precisely those where $N_+-N_-=m$.

\begin{figure}[ht] 
%\centerline{\epsfxsize=4.00in\ \epsfbox{nu_fig.ps}}
\centerline{\epsfxsize=3.20in\ \epsfbox{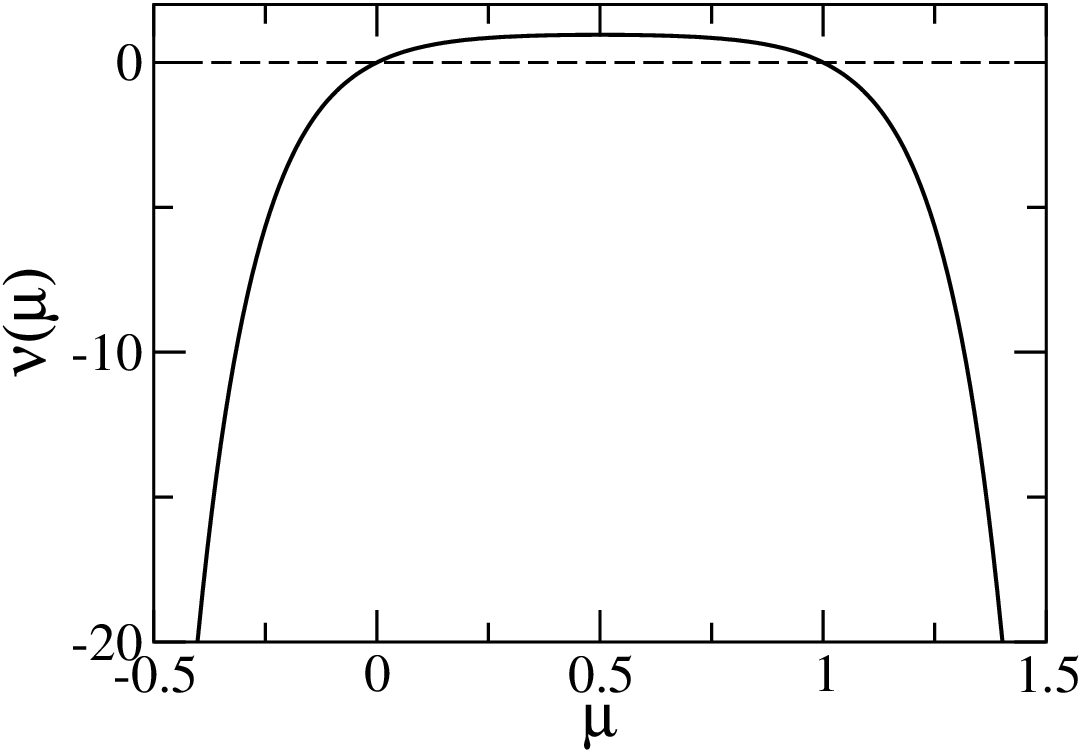}}
\caption{The eigenvalue $\nu(\mu)$, plotted as a function of $\mu$, for $\t=0.001$.  The top of the curve becomes flatter
as $\t\to0$.
}\label{fig2}
\end{figure}

Upon making the substitution $z=e^{-\mu \, ds}=(\wm/\wp)^\mu$, the generating function assumes its usual interpretation:
\begin{equation}
\label{p}
p(e^{-\mu \, ds})=\av{\exp(-\mu s)}=\left(\wo+\wp^{1-\mu}\wm^\mu+\wp^\mu\wm^{1-\mu}\right)^M\,.
\end{equation}
The time evolution of this generating function is described by a linear operator whose lowest eigenvalue, $\nu(\mu)$, 
allows one to compute quantities of interest \cite{Leb99,Meh08,Dor11}.  For now, we ignore the linear operator itself,
since we can obtain $\nu$ directly, from
\[
\nu(\mu)=\lim_{T\to\infty}\left[-\frac{1}{T}\ln\av{\exp(-\mu s)}\right]\,,
\]
where $T=M\tau$ is the total time length of each trajectory.  We thus obtain
\begin{equation}
\label{nu}
\nu(\mu)=-\t^{-1}\ln\left(\wo+\wp^{1-\mu}\wm^\mu+\wp^\mu\wm^{1-\mu}\right)\,.
\end{equation}
Interestingly, in our case this limit is achieved for {\em any} value of $M$.  This helps us discuss  $T\to\infty$ ($M\to\infty$), even as 
$\tau\to0$, for we can take the two limits independently. The fact that $\nu(\mu)=\nu(1-\mu)$ 
is a manifestation of the detailed fluctuation theorem \cite{Eva93,Gal95,Kur98,Leb99}.  The eigenvalue $\nu(\mu)$ is plotted in Fig.~\ref{fig2}, for the case of $\t=0.001$.
Using the low-order approximations for the $\w$'s we get
\begin{equation}
\nu(\mu)\approx 1-(\tau/2)^\mu-(\tau/2)^{1-\mu}\,,
\end{equation}
which compares very nicely with (\ref{nu}) when $\t\to0$.

%%%%%%%%%%%%%%%%%%%%%%%%%%%%%%%%%%%%%%%%%%%FIG 2.%%%%%%%%%%%%%%%%%%%%%%%%%%%%%%%%%%%%%%%%%%%%%%%%%%%%%%
%%%%%%%%%%%%%%%%%%%%%%%%%%%%%%%%%%%%%%%%%%%FIG 2.%%%%%%%%%%%%%%%%%%%%%%%%%%%%%%%%%%%%%%%%%%%%%%%%%%%%%%

The mean entropy production rate can  now be derived from $\nu(\mu)$:
\begin{equation}
\av{\dot s} = d\nu/d\mu|_{\mu=0}=\t^{-1}(\wp-\wm)\ln(\wp/\wm)\approx\ln(2/\t)\,,
\end{equation}
where we have used $\wo+\wp+\wm=1$, and the last expression is the dominant behavior as $\t\to0$.  The fact that the approximate limit is the same as the entropy produced in a single forward transition is in agreement with the notion that backward steps are exceedingly rare as $\t\to0$ and do not contribute to the average.  

The mean entropy production may also be computed from 
\begin{equation}
\av{\dot s}=\tau^{-1}\sum_{X,Y}\r(X)\w(X,Y)\ln\frac{\w(X,Y)}{\w(Y,X)}\,.
\end{equation}
In general, the sum is dominated by the states $P,Q$ yielding the fastest diverging $\w(P,Q)/\w(Q,P)$ ratio, as $\tau\to0$.  The
dominant contribution comes from an irreversible transition, $P\to Q$ and $Q\not\to P$, since 
$\w(Q,P)\to0$, as $\tau\to0$, in that case.  It is in this sense that our model is canonical, for it suffices to focus on the effect of a single (dominant) irreversible transition, and ours is the smallest model that accomplishes that.

The fluctuation function, $\chi(\sigma)$, of the {\em scaled} entropy,  $\sigma=s/\av{\dot s}T$, is derived from an extremum of the Legendre transform of $\nu$:
\begin{equation}
\chi(\sigma)=\max_\mu\{\nu(\mu)-\av{\dot s}\sigma\mu\}\,.
\end{equation}
It is possible to obtain a full analytic derivation of $\chi(\nu)$ for our simple model, but this results in cumbersome expressions.  Instead, we illustrate the technique for the limit of small $\tau$.  The two derivations yield virtually
indistinguishable curves, for $\tau\lesssim0.001$, while  more insight is gained from the simpler  approximation.

We begin by rewriting (the approximate) $\nu(\mu)$ as 
\[
\nu(\mu)=1-x-\frac{\t }{2}x^{-1}\,;\qquad x\equiv(\t/2)^\mu\,,
\]
and find $\mu_*$ that maximizes $\nu(\mu)-\av{\dot s}\sigma\mu$, using the approximate limit $\av{\dot s}=\ln(2/\t)$;
\[
x_*=\frac{\sigma+\sqrt{\sigma^2+2\t}}{2}\,,\qquad \mu_*=\frac{\ln x_*}{\ln(\t/2)}\,.
\]
(The other root of the quadratic equation for $x$ yields unphysical, complex values.)  Finally, putting $x=x_*$ and 
$\mu=\mu_*$ in $\nu(\mu)-\av{\dot s}\sigma\mu$, we obtain
\begin{equation}
\label{chi_approx}
\chi(\sigma)=1-\sqrt{\sigma^2+2\t}+\sigma\ln\left(\frac{\sigma+\sqrt{\sigma^2+2\t}}{2}\right)\,.
\end{equation}
It is easy to check that this satisfies the symmetry relation $\chi(-\sigma)=\chi(\sigma)+\av{\dot s}\sigma$, yet another manifestation of the detailed fluctuation theorem.

\begin{figure}[h] 
\centerline{\epsfxsize=3.20in\ \epsfbox{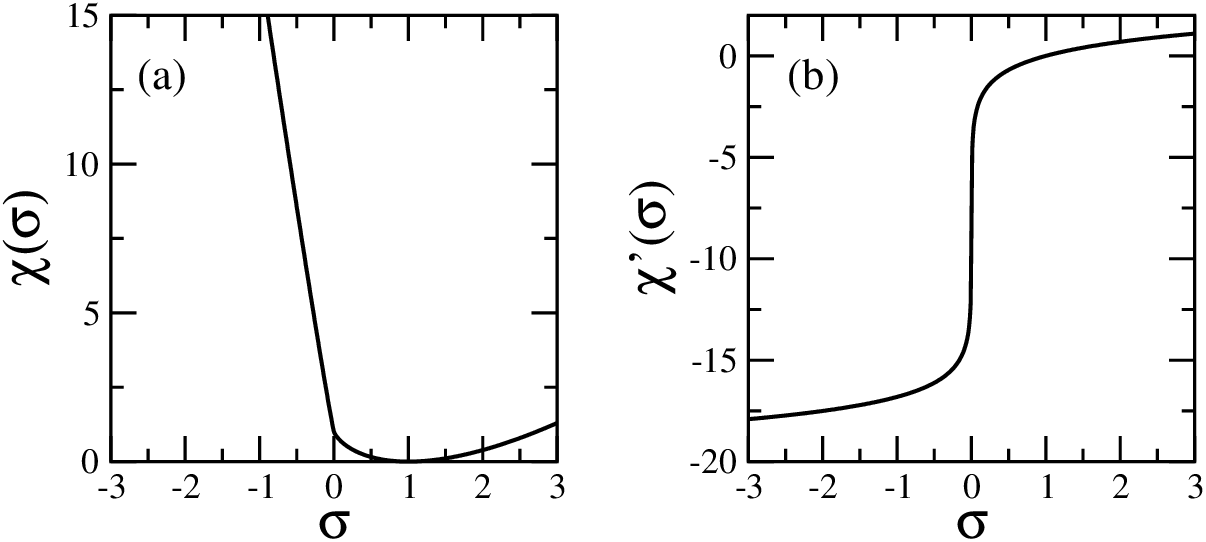}}
\caption{The large deviation function (a), and its derivative~(b), as a function of $\sigma$, for $\t=10^{-7}$.  The kink is more pronounced the smaller the value of $\t$.
}\label{fig3}
\end{figure}

The limiting form of $\chi(\sigma)$ is universal:
\begin{equation}
\label{chi_t_0}
\chi(\sigma)\longrightarrow 1-\sigma+\sigma\ln\sigma,\quad {\rm as\ }\t\to0\,;\qquad\sigma>0\,,
\end{equation}
and $\chi(\sigma)\to\infty$ for $\sigma<0$, as $\t\to0$.
The origin of the  kink \cite{Meh08} in $\chi(\sigma)$ resides in $\sqrt{\sigma^2+2\t}$, Eq.~(\ref{chi_approx}), which tends to $|\sigma|$ as $\t\to0$.  Moreover, at the same limit, the logarithmic term diverges for $\sigma<0$, but not for $\sigma>0$.  The kink can be best explored through the derivatives of $\chi(\sigma)$ \cite{Dor11}:
\begin{equation}
\begin{split}
&\chi'(\sigma) = \ln\left(\frac{\sigma+\sqrt{\sigma^2+2\t}}{2}\right) \,,\\
&\chi''(\sigma) = \frac{1}{\sqrt{\sigma^2+2\t}}\longrightarrow|\sigma|^{-1} \,,
\end{split}
\end{equation}
as $\t\to0$.
Note the existence of the limit $\tau\to0$ for $\chi''(\sigma)$ for all $\sigma\neq0$.   For finite $\t$ 
the magnitude of the apparent jump in $\chi'$ is found to be of order $\chi'(1)-\chi'(-1)\to\ln(2/\t)$.
The large deviation function $\chi(\sigma)$ and its derivative are plotted in Fig.~\ref{fig3}.

%%%%%%%%%%%%%%%%%%%%%%%%%%%%%%%%%%%%%%%%%%%FIG 3.%%%%%%%%%%%%%%%%%%%%%%%%%%%%%%%%%%%%%%%%%%%%%%%%%%%%%%
%%%%%%%%%%%%%%%%%%%%%%%%%%%%%%%%%%%%%%%%%%%FIG 3.%%%%%%%%%%%%%%%%%%%%%%%%%%%%%%%%%%%%%%%%%%%%%%%%%%%%%%

For our simple model, we were able to find the generating function~(\ref{p}) by inspection.  For other systems, 
in general, it can be expressed as a matrix product,
\begin{equation}
\av{e^{-\mu s}}={\bf u}\,{\bf R}^M{\bf v};\quad {\bf R}_{X,Y}=\w(X,Y)^{1-\mu}\w(Y,X)^{\mu}\,,
\end{equation}
where ${\bf u}=(\rho(A),\rho(B),\dots)$ and ${\bf v}$ is a column vector of ones.  Then, for $T\to\infty$,
\begin{equation}
\nu(\mu)=-\t^{-1}\,{\ln \lambda(\mu)}\,,
\end{equation}
where $\lambda(\mu)$ is the largest eigenvalue of ${\bf R}$.

{\em $N$-state ring.---}
It is easy to generalize the foregoing results to an $N$-state ring: $A_1\to A_2,\,A_2\to A_3,\dots, A_N\to A_1$, ($N\geq3$),
where all rates are 1.  The key ingredient arises from the fact that the forward transition probability (after time $\t$), from $A_k\to A_{k+1}$, is then $\approx \tau$, while the ``forbidden" transition probability, for $A_k\to A_{k-1}$, is
$\tau^{N-1}/(N-1)!\equiv\delta_N\tau$.  All of the results valid for $N=3$ can be then extended to general $N$, expressed as a function of $\delta_N$.
In particular, 
\begin{equation}
\nu(\mu)=1-\delta_N^\mu-\delta_N^{1-\mu}\,,
\end{equation}
from which follows
\begin{eqnarray}
&&\av{\dot s} = \ln(1/\delta_N)\,,\\
&&\chi(\sigma) = 1-\sqrt{\sigma^2+4\delta_N}\\ \nonumber
&& \hspace*{1.4cm} +\sigma\ln\left(\frac{\sigma+\sqrt{\sigma^2+4\delta_N}}{2}\right)\,,\\
&&\chi'(\sigma)|_{\t\to0} = -\ln\left(\frac{\sigma+\sqrt{\sigma^2+4\delta_N}}{2}\right)\,.
\end{eqnarray}
The results $\chi(\sigma>0)|_{\t\to0}=1-\sigma+\sigma\ln\sigma$, and $\chi''(\sigma)|_{\t\to0}=1/|\sigma|$,  are universal.

{\em Event sampling.---}
The $N$-state ring can be analyzed also with event sampling, only that then one must postulate \cite{Dor11} a small back
reaction rate $\epsilon$ for the ``forbidden" transitions $A_k\to A_{k-1}$.  It is easy to show that 
\begin{equation}
\nu(\mu)=1+\epsilon-\epsilon^\mu-\epsilon^{1-\mu}\,,
\end{equation}
for all $N\geq3$.
Thus, the results from event sampling agree with those of interval sampling, in the limit of $\t\to0$, provided that one sets
$\epsilon=\delta_N=\t^{N-2}/(N-1)!$ (for the $N$-ring).  This physical meaning of the small rate $\epsilon$ is new to our work---indeed, for event sampling there is no coherent prescription on how to choose independent $\epsilon$'s for the various irreversible transitions.

{\em Conclusion.---} In this letter we have proposed the use of interval sampling, a novel technique for studying entropy production in nonequilibrium steady states.  Most importantly, interval sampling allows direct analysis of systems with microscopic irreversibility, and is more easily implemented in experiments.   We then focused on the smallest 
model possessing irreversibility --- the three-state loop --- and argued that it may serve as a canonical example for
systems with microscopic irreversibility, such as driven granular systems, in general.  In this way, we were able to identify universal features of  entropy production, including its large deviation function and the kink at zero entropy production, which is now seen to clearly arise from the irreversibility alone.

%In this letter we have shown that fluctuation theorems involving entropy production are naturally obtained
%when considering interval-driven paths. As these paths are easily realized in experiments, this opens new types of systems, as for
%example driven granular systems, to a study involving fluctuation theorems. In addition, we analyzed in detail a model where all
%relevant quantities can be obtained in closed analytical form. This is arguably the simplest model that can be studied in this
%context and that does yield non-trivial results. Thus, we are able to investigate the kink in the entropy function, which is a
%generic feature of the steady-state entropy production. Finally, in the irreversible limit, the large deviation function
%approaches a universal form.

This work was supported by the US National
Science Foundation through DMR-0904999 as well as by the National Research Fund, Luxembourg, and cofunded under the Marie Curie Actions of 
the European Commission (FP7-COFUND).

\end{document}